\documentclass[prl,aps,floats,floatfix,superscriptaddress,preprintnumbers,showpacs,twocolumn]{revtex4}
\usepackage{amsfonts,amsmath,epsfig,amssymb,latexsym,eucal,color,graphicx,array,subfigure,bm}

\definecolor{red  }{rgb}{1,0,0}
\definecolor{blue }{rgb}{0,0,1}
\definecolor{green}{rgb}{0,1,0}

\usepackage{graphicx,amssymb,amsmath,bm,latexsym}
\newcommand{\nt}{n_{\scriptscriptstyle T}}
\newcommand{\ns}{n_{\scriptscriptstyle S}}
\newcommand{\reh}{\tilde R_{\rm {\scriptscriptstyle H}}}
\newcommand{\rch}{\tilde R_{\rm {\scriptscriptstyle C}}}

\newcommand{\vect}[1]{\!\!\!\mbox{ \boldmath $#1$}}

\begin{document}
\title{Black Holes in an Expanding  Universe}

\author{Gary W. {\sc Gibbons}}
\address{DAMTP, Centre for Mathematical Sciences, Cambridge University, 
Wilberforce Road, Cambridge CB3 OWA, UK}

\author{Kei-ichi {\sc Maeda}}
\address{Department of Physics \& RISE, Waseda University, Okubo 3-4-1, 
Shinjuku, Tokyo 169-8555, Japan}


\begin{abstract}
An exact solution representing  black holes in an  expanding universe is found.
The black holes are  maximally charged and the universe is 
expanding with arbitrary equation of state ($P=w \rho$ with $-1\leq
 \forall w \leq 1$).
It is an exact solution of  the Einstein-scalar-Maxwell system, 
in which we have two Maxwell-type U(1) fields 
coupled to the scalar field.
The potential of the scalar field is
an exponential.
We find a regular horizon, which depends on one parameter 
(the ratio of the energy density of U(1) fields to that of the scalar field).
The horizon is static because of the balance on the horizon
between gravitational attractive 
force and U(1) repulsive force acting on the scalar field.
We also calculate the black hole temperature.
\end{abstract}
\maketitle

Black holes play a central role in astrophysics as well as string theory, 
and may  possibly be created at the LHC \cite{TeV_BH} . Not surprisingly, 
they have been studied intensively over the past 35 years. Important progress 
has been made in understanding the Hawking process \cite{Hawking1974} and the
states responsible for black hole entropy at the microscopic 
level\cite{Strominger:1996sh} .  However many
problems remain unresolved: does cosmic censorship \cite{Penrose} hold?,
 what happens when black holes collide?,
how does accretion of matter affect the thermodynamics  of black holes? 
and  how does it affect the growth of black holes?
To answer these dynamical questions one needs time-dependent solutions of 
the Einstein equations containing black holes.
 In this letter we shall focus on solutions representing black
 holes in a 
background
 FLRW universe.

There have been many previous  attempts to 
obtain black holes 
embedded in the Friedmann-Lema\^itre-Robertson-Walker (FLRW) universe. 
The Einstein-Straus model 
 is perhaps  the simplest one~\cite{Einstein:1945id}.
It is a patchwork of  
 Schwarzschild black holes with an FLRW universe. 
However these black-holes are time symmetric,  
and so they do not describe a dynamical black hole
in the Universe. 

One well  known black hole candidate in FLRW universe 
is the McVittie
solution~\cite{Mcvittie1933},  which is a spherically symmetric, 
time-dependent solution of the Einstein equations.
The solution approaches an FLRW
universe at ``infinity,'' and looks like a black hole near the ``horizon''.
However, as shown in~\cite{Nolan}, the McVittie solution 
is disqualified as a black hole (or a point mass singularity) 
in the FLRW universe.
Recently,
Sultana and Dyer constructed a more sophisticated black hole solution 
in a dynamical background by a conformal technique~\cite{SultanaDyer}. 
The matter content is  null dust and  ordinary dust.
The solution tends to an Einstein-de Sitter spacetime asymptotically.  
This model, however, violates the dominant energy
conditions. 

Assuming self-similarity, we can show that 
a regular black hole may exist only in an accelerating universe,
but this requires  numerical study\cite{MHC}.
The analytic solution found by Carr and Hawking
 describes a self-similar spacetime with a 
regular black hole but it approaches asymptotically
a ``quasi" FLRW spacetime which has a deficit angle, 
but not an exact flat FLRW spacetime\cite{CH}.
There are also  discussions of
``dark energy" accretion onto a black hole in  a 
  universe.
\cite{accretion_BH_universe}.

If a cosmological constant  is present,
we have the Schwarzschild-de Sitter (SdS) and 
Reissner-Nordstr\"{o}m-de Sitter (RNdS) solutions~\cite{SdS,Carter}.
Although these spacetimes are static,
they  may be converted  by a coordinate transformation
into  the form of a black hole in an exponentially expanding 
Universe~\cite{BH}. 
Multi-black hole solutions
in a  de Sitter Universe were  found 
by use of extremely charged black holes
 and  their collision discussed~\cite{KT,BHKT}.
This Kastor-Traschen (KT) 
solution is a time-dependent generalization of 
the Majumdar-Papapetrou solution, which describes 
extremely charged RN black holes~\cite{Hartle:1972ya}. 
Similar solutions 
were given in \cite{BC,Sh}.

Another time-dependent  cosmological black hole system  was found
from the compactification of
intersecting brane solution in higher dimensional unified theory\cite{MOU}.
As clarified   in \cite{MN} the 
global picture of dynamical solution 
 describes a multi-black hole system in the expanding Universe
filled by ``stiff matter". We shall call it the MOU solution.

Here we generalize these two solutions and present an exact solution
 describing  a cosmological  multi-black hole system  
with an arbitrary power law expansion.
This is a solution of  general relativity with a scalar field with
 an exponential potential and two Maxwell-type U(1) fields coupled 
to the scalar field.
The solution has regular event horizons, approaches asymptotically
an exact flat FLRW spacetime without a deficit angle,
and 
no singularity exists outside the horizons.


The above 
known
 solutions take the following form:
\begin{eqnarray}
ds^2=-\bar U^{-2}d\bar t^2+a^2(\bar t) \bar U^2 d\,\vect{r}^2
\label{metric0}
\,.
\end{eqnarray}
The KT solution with $N$-black holes
 located at 
the coordinate position $\vect{r}_{(i)}$  ($i=1, \cdots, N$)
 is given by
\begin{eqnarray}
\displaystyle{
\bar U=1+\sum_{i=1}^N {Q_{(i)}\over a|\,\vect{r}-\vect{r}_{(i)}|}}
\,,
\end{eqnarray}
where $Q_{(i)}$ is the charge of the $i$-th black hole,
  and the scale factor of the background Universe is given by 
$a(\bar t) \propto \exp(H_0 \bar t)$ ($H_0$: constant).
It is asymptotically de Sitter spacetime.
The MOU solution discussed in \cite{MOU,MN} is 
given by
$$
\bar U=\left[1+\sum_{i=1}^{N}{Q_{T(i)}
\over a^4|\,\vect{r}-\vect{r}_{(i)}|}\right]^{1/4}
\left[1+\sum_{i=1}^N{Q_{S(i)}\over |\,\vect{r}-\vect{r}_{(i)}|}\right]^{3/4}
\hskip -1em 
\,,
$$
where $Q_{T(i)}$ and  $Q_{S(i)}$ are the conserved charges of 
time-dependent and static branes, respectively, and 
the scale factor is given by 
$a= (\bar t/\bar t_0)^{1/3}$,
which also holds  for an expanding universe with
stiff matter.

By changing the time coordinate,
 these solutions can be rewritten in the form of a brane system,
 discussed in detail in  \cite{MOU}, as
\begin{eqnarray}
ds^2=-U^{-2}dt^2+ U^2 d\,\vect{r}^2
\label{metric1}
\,,
\end{eqnarray}
where
\begin{eqnarray}
&&
U=H_T^{\nt/4}H_S^{\ns/4}
\label{metric_U}
\,,
\\
&&
~
H_T={t\over t_0}+\sum_{i=1}^{N}{Q_{T(i)}\over |\,\vect{r}-\vect{r}_{(i)}|}
\,,~~
\label{harmonics_T}
\\
&&
~
H_S
=1+\sum_{i=1}^N{Q_{S(i)}\over |\,\vect{r}-\vect{r}_{(i)}|}
\,.~~~~
\label{harmonics_S}
\end{eqnarray}
Here $n_T$ and $n_S$ are appropriate non-negative real numbers
with the constraint $\nt+\ns=4$, and
$t_0$ is a constant.
The transformation of the time coordinate is
given by 
${t/t_0}=a^{4/\nt}(\bar t)$, where
$t_0$ is fixed as
$t_0=H_0^{-1}$ for the KT solution
and $3\bar t_0/4$
 for the MOU solution,
respectively.
Setting $\nt=4$, we find
 the KT solution, 
while the case with $\nt=1$ 
corresponds to the MOU solution.

Assuming the metric form (\ref{metric1}) $\sim$ (\ref{harmonics_S}),
if we take an arbitrary real value for $\nt$ (or $\ns$), we find that 
the scale factor $a$ in the form of (\ref{metric0})
is given by any power function,
i.e., $a\propto \bar t^{\,p}$, where
$p=\nt/\ns$. (We regard $p=\infty$ as an exponential expansion.)

When the universe expands with an arbitrary power by a scalar field,
one needs an exponential-type potential\cite{power_law}.
In fact, if we have the universe filled by a scalar field
with the potential 
\begin{eqnarray}
V=V_0 \exp (-\alpha\kappa\Phi)
\label{exp_pot}
\,,
\end{eqnarray}
the scale factor increases as
$a\propto \bar t^{\,p}$ with $p=2/\alpha^2$,
where 
$\kappa^2=8\pi G_N$ is the gravitational constant.
Thus  we should choose 
$\alpha=\sqrt{2/p}=\sqrt{2\ns/\nt}$.

We shall therefore adopt the following action:
\begin{eqnarray}
S&=&\int d^4x\sqrt{-g}\Bigl[
{1\over 2\kappa^2}R
-{1\over 2}g^{\mu\nu}(\partial_\mu\Phi)
(\partial_\nu\Phi)
\nonumber \\
&
-&
V(\Phi)
-{1\over 16\pi}\sum_{I=S,T}  n_I e^{\lambda_I\kappa\Phi}
(F_{\mu\nu}^{(I)})^2\Bigr]
\,,
~~~
\label{action}
\end{eqnarray}
where 
$g_{\mu\nu}$ is a spacetime metric,
$\Phi$ is a scalar field with 
the potential $V(\Phi)$
 given by (\ref{exp_pot}),
and   $F_{\mu\nu}^{(I)} (I=S,T)$
are two Maxwell-type U(1) fields, which couple 
to the scalar field with the coupling
constants $\lambda_I$. The vector potentials
are given by  $A_{\mu}^{(I)}$, and 
$n_I$ are their degeneracy factors.

The metric (\ref{metric1}) with
(\ref{metric_U}) plus
\begin{eqnarray}
&&
\kappa\Phi={1\over 2}\sqrt{\nt\ns/2}
\,\ln \left({H_T/H_S}\right)
\,,
\\
&&
\kappa A^{(T)}_{~t}=\sqrt{2\pi}\,H_T^{-1}
\,,~~
\kappa A^{(S)}_{~t}=\sqrt{2\pi}\,H_S^{-1}
\,,~~~
\end{eqnarray}
with
(\ref{harmonics_T}), (\ref{harmonics_S})
and
$
\kappa^2 V_0 t_0^2={\nt(\nt-1)/4}
$
is really an exact solution of the 
system (\ref{action}), if we assume 
\begin{eqnarray}
\alpha=
\lambda_T=\sqrt{2\ns/ \nt}
\,,
~~
\lambda_S=-\sqrt{2\nt/\ns}
\,,
\end{eqnarray}
and $\nt+\ns=4$
\cite{footnote}.
For $\nt=4$ and $\nt=1$, we recover the KT and MOU solutions, respectively.

The above  solution with arbitrary $\nt$
gives a multi-black hole system in an expanding universe
for  which the scale factor and effective equation of state
are given by
 $a\propto \bar t^{\,p}$ with $p=\nt/\ns$,
and $P=w\rho$ with $w={2\ns\over 3\nt}-1$,
respectively.
Note that $w$ takes an arbitrary value in the range of
$-1\leq w\leq 1$, corresponding to the value of $1\leq\nt \leq 4$.

We summarize some typical solutions in Table \ref{table1}.

In order to discuss the spacetime found here in detail,
in what follow, we consider a single black hole system.
For simplicity, we assume that two charges are equal, i.e.,
$Q_T=Q_S=:Q$.
We shall rewrite the metric (\ref {metric1})
as
\begin{eqnarray}
d\tilde s^2&=&-\tau^2 U^{-2} d\tilde t^2+U^{2}
\left(d\tilde r^2+\tilde r^2d\Omega_2^2\right)
\label{metric2}
\end{eqnarray}
with (\ref{metric_U}) and 
$
H_T=\tilde t+ \tilde r^{-1}$, $H_S
=1+\tilde r^{-1}$,
where $\tau=t_0/Q$, $d \tilde s^2=ds^2/Q^2$, $\tilde r=r/Q$, and 
$\tilde t=t/t_0$
are dimensionless variables.
The metric (\ref{metric2}) depends on only one parameter
$\tau$, whose  physical meaning is given as follows:
The energy density of the scalar field is uniform at $t=t_0$, which 
is given by 
$\rho_\Phi(t_0)=3\nt^2/16t_0^2$.
While the total density of the U(1) fields 
is evaluated on the horizon for the static extreme
RN black hole with the charge $Q$
as $\rho_{\rm {\scriptscriptstyle U(1)}}|_{\scriptscriptstyle H}=1/Q^2$.
For the time-dependent black hole, 
$\rho_\Phi|_{\scriptscriptstyle H}$ and
$\rho_{\rm {\scriptscriptstyle U(1)}}|_{\scriptscriptstyle H}$ are
different from the above values, but their orders 
of magnitude are still the same. Thus
we can  claim that $\tau$ is related to
the ratio of two energy densities at the horizon as
$
\tau^2\sim
\rho_{\rm  {\scriptscriptstyle U(1)}}
/ \rho_\Phi|_{\scriptscriptstyle H}
$
(see \cite{MN,MN2}). 
The limit of $\tau\rightarrow \infty$
gives the static extreme
RN black hole.

\vskip -.7cm
\begin{widetext}
\begin{center}
\begin{table}[h]
\footnotesize{
\begin{tabular}{|c||c|c||c|c|c||cl|c|c|}
\hline
\raisebox{-.6em}{Type}&
\raisebox{-.6em}{$\nt$}&\raisebox{-.6em}{$\ns$}&
\multicolumn{3}{|c||}{\raisebox{-.3em}{The coupling constants}}&
\multicolumn{3}{|c|}{\raisebox{-.3em}{Models of the Universe}}&
\raisebox{-.6em}{$
\kappa^2 V_0 t_0^2 $}
\\[-.2em] 
\cline{4-9}
&&&$\alpha$
&\raisebox{.1em}{$\lambda_T$}
&$\lambda_S$&\raisebox{.1em}{$p$}
&{(expansion law)}&{$w$}&
\\ 
\hline
\hline
&0&4&$\infty$& $\infty$& $0$ & ~$0$
&({\bf static}) &$0$ & 0
\\ \cline{2-10}\cline{2-10}
\raisebox{-.6em}{I}
&1&3&$\sqrt{6}$&$\sqrt{6}$&$-{\sqrt{6}/ 3}$ & ~${1/ 3}$  
 &({\bf stiff matter}) &1& 0
\\[-.3em] \cline{2-10}
&${4/ 3}$&${8/ 3}$&2& 2& $-1$ & ~${1/ 2}$&({\bf radiation}) 
&${1/ 3}$ & ${1/ 9}$ \\ \cline{2-10}
&${8/ 5}$&${12/ 5}$&$\sqrt{3}$ & $\sqrt{3}$ &
 $-{2/ \sqrt{3}}$& ~${2/ 3}$
&({\bf dust}) &0& 
${6/ 25}$ \\ \hline
II&2&2&$\sqrt{2}$& $\sqrt{2}$& $-\sqrt{2}$&~$1$&({\bf Milne}) &
$-{1/ 3}$& 
${1/ 2}$\\ \hline
\raisebox{-.6em}{III}
&3&1&${~\sqrt{6}/ 3~}$ & ${~\sqrt{6}/ 3~}$ & $-\sqrt{6}$&
 ~$3$ &({\bf quintessence})
&$-{7/ 9}$& ${3/ 2}$
\\[-.3em] \cline{2-10} 
&4&0&0&  0 &$-\infty$& $\infty$ &({\bf de Sitter})&$-1$& 3
\\
\hline
\end{tabular} }
\caption{Some values of the
typical parameters of a black hole system in a universe,
for  which  expansion law and the equation of state are  given by
$a\propto \bar t^{\,p}$ and $P=w\rho$, respectively.
}
\label{table1}
\end{table}
\end{center}
\vskip -2.5em
\end{widetext}
\vskip -.7cm

The circumference radius $R=Q \tilde R$,
which is a geometrically invariant variable,
 is given by
\begin{eqnarray}
\tilde R=\tilde r U=\left(1+\tilde t\tilde r
\right)^{\nt\over 4}
\left(1+\tilde r\right)^{\ns\over 4}
\label{circumference}
\,.
\end{eqnarray}

The curvature singularity appears at $\tilde r=-1$ and $\tilde r
=-1/\tilde t$, where $\tilde R$ vanishes.
Analyzing the behaviour of trapping horizons in the limit
of $\tilde r\rightarrow 0$ 
and near horizon geometry as in \cite{MN},
we find the horizon radius ($\tilde R_+$
or $\tilde R_-$) which satisfies
the following equation:
\begin{eqnarray}
\tau\,(
\tilde R_+^{4\over \nt}-1\,)
= \tilde R_+^2
\,,~~
\tau\,(
\tilde R_-^{4\over \nt}-1\,)
=- \tilde R_-^2
\,.
~~~
\label{horizon_pm}
\end{eqnarray}

The spacetimes are 
classified by their causal structure  into three types:
Type I ($\nt<2 $),  
Type II ($\nt=2 $), and 
Type III ($\nt>2 $).

\begin{figure}[h]
\begin{center}
\includegraphics[width=3.5cm]{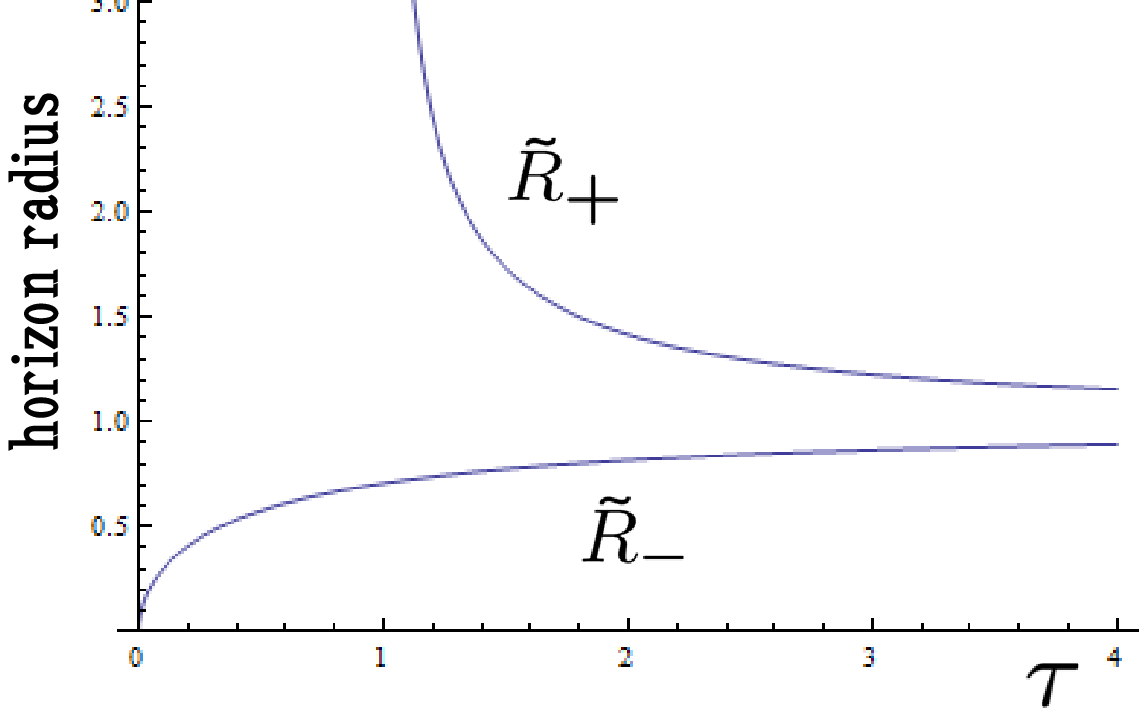}
\includegraphics[width=3.5cm]{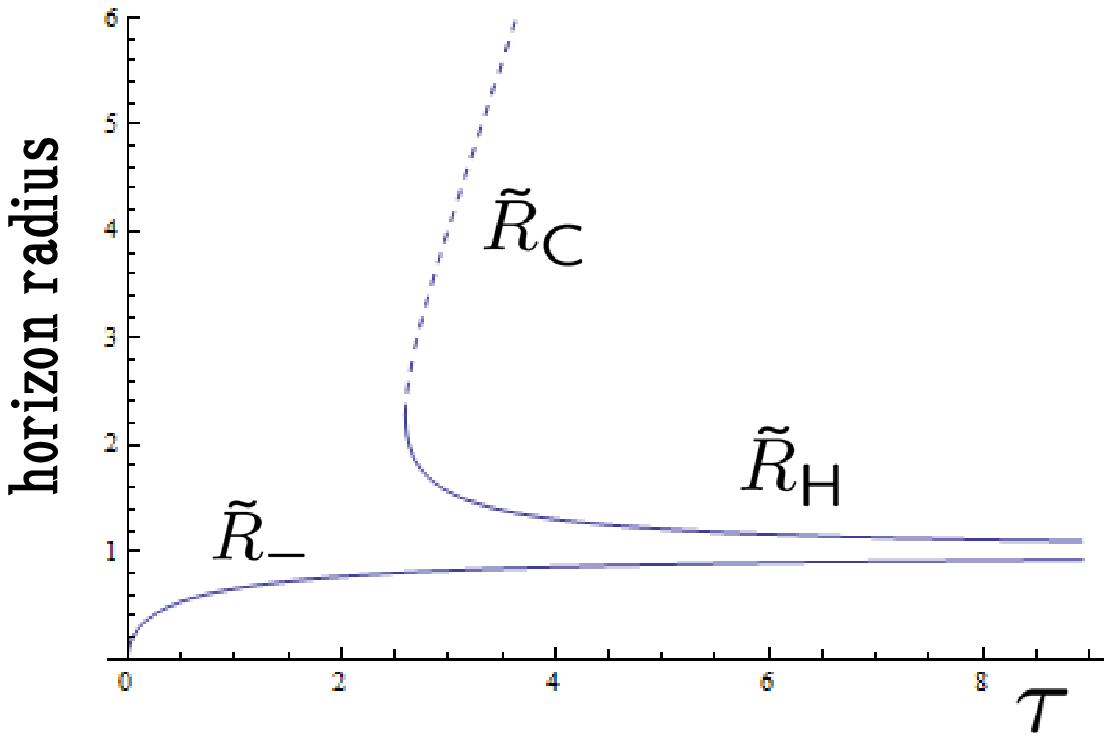}
\\
\hskip .2cm (a) $\nt=1$
\hskip 1.8cm (b) $\nt=3$
\caption{
Two horizon radii ($\tilde R_-$ 
and $\tilde R_+$ or $\reh$) for $\nt=1$ 
and $3$.
$\reh$ and $\rch$, which is not  a cosmological horizon,
 degenerate at 
$\tau_{\rm cr}=3\sqrt{3}/2$.}
\label{fig:horizon}
\end{center}
\vskip -2em
\end{figure}

In Type I, there are two horizons, $\tilde R_+$ and $\tilde R_-$, 
which are
the roots of Eqs. (\ref{horizon_pm}).
Since $\tau>0$, we find $\tilde R_+>1>\tilde R_->0$.
We show  the horizon radii 
in terms of $\tau$ in Fig. \ref{fig:horizon} (a).
For Type II, if $\tau>1$ there are two horizons,
 $\tilde R_+=\sqrt{\tau/(\tau-1)}$ and $\tilde R_-=\sqrt{\tau/(\tau+1)}$,
but if $\tau\leq 1$, we find only one horizon, 
$\tilde R_-$. 
In Type III, if $\tau>\tau_{\rm cr}$, we find 
two roots $\tilde R_{+,1}$ and $\tilde R_{+,2}$
($\tilde R_{+,1}<\tilde R_{+,2}$) 
for the equation for $\tilde R_{+}$ (\ref{horizon_pm}), 
where $\tau_{\rm cr}=\nt^{\nt/2}
(\nt-2)^{-(\nt-2)/2}/2$. 
From the detail analysis of spacetime structure,
we find there exist two horizons;
$\tilde R_-$ and $\reh=\tilde R_{+,1}$
($\tilde R_-<1<\reh$), but
 $\rch
=\tilde R_{+,2}$
is not a cosmological horizon except for 
the KT solution\cite{MN2}.
The cosmological horizon turns out to be
time-dependent just as that in an accelerating universe
\cite{HKS}.
On the other hand, if $\tau<\tau_{\rm cr}$, we find one horizon,
$\tilde R_-$. If $\tau=\tau_{\rm cr}$, there are two horizons,
 $\tilde R_-$ and $\tilde R_{+,1}=\tilde R_{+,2}$ 
(degenerate) (see Fig. \ref{fig:horizon} (b)).
As we see from Table \ref{table1}, 
Type I and III correspond to a decelerating
and accelerating universes, respectively.

\begin{figure}[h]
\begin{flushleft}
\includegraphics[width=3.5cm]{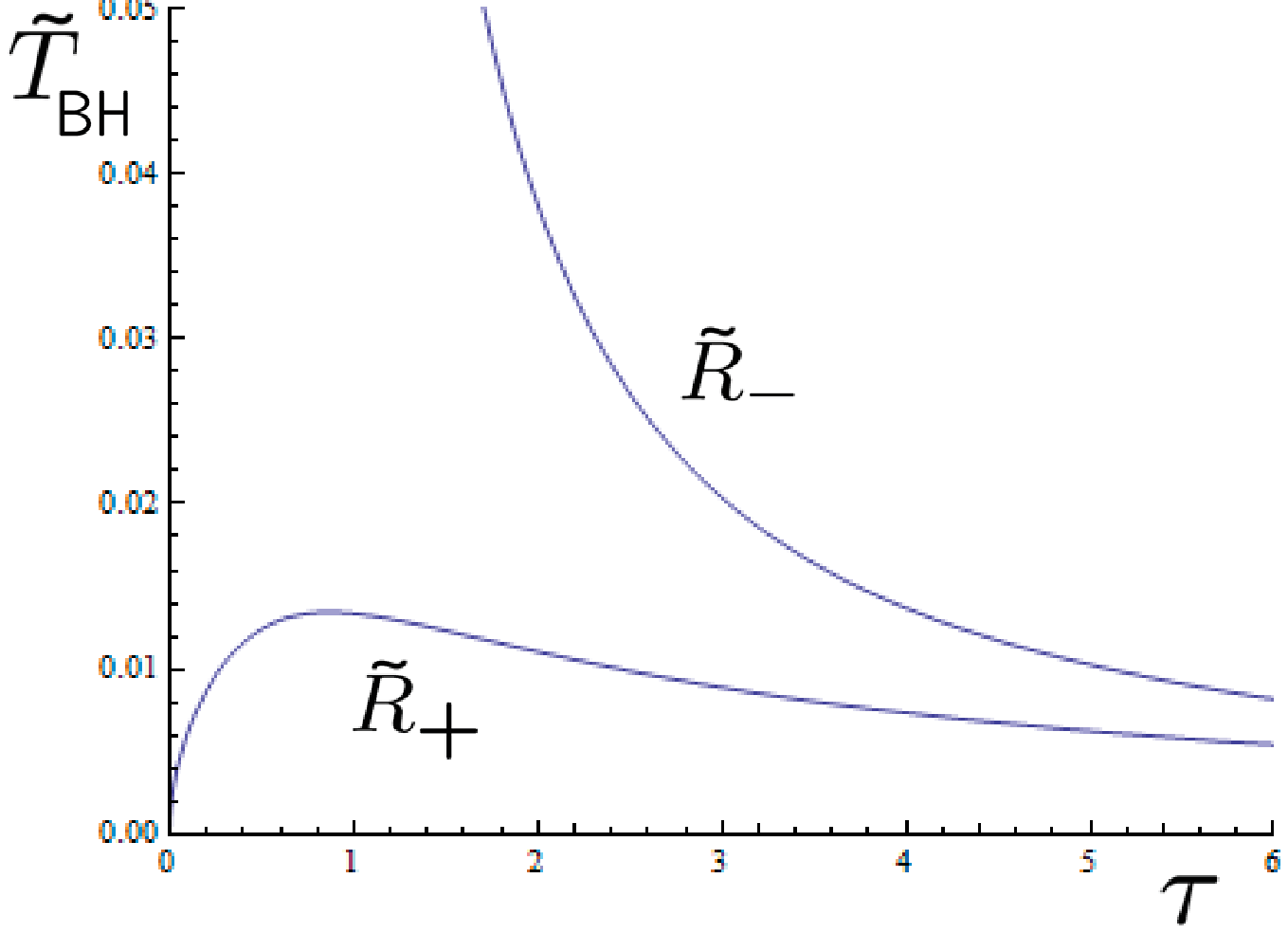}
\includegraphics[width=3.5cm]{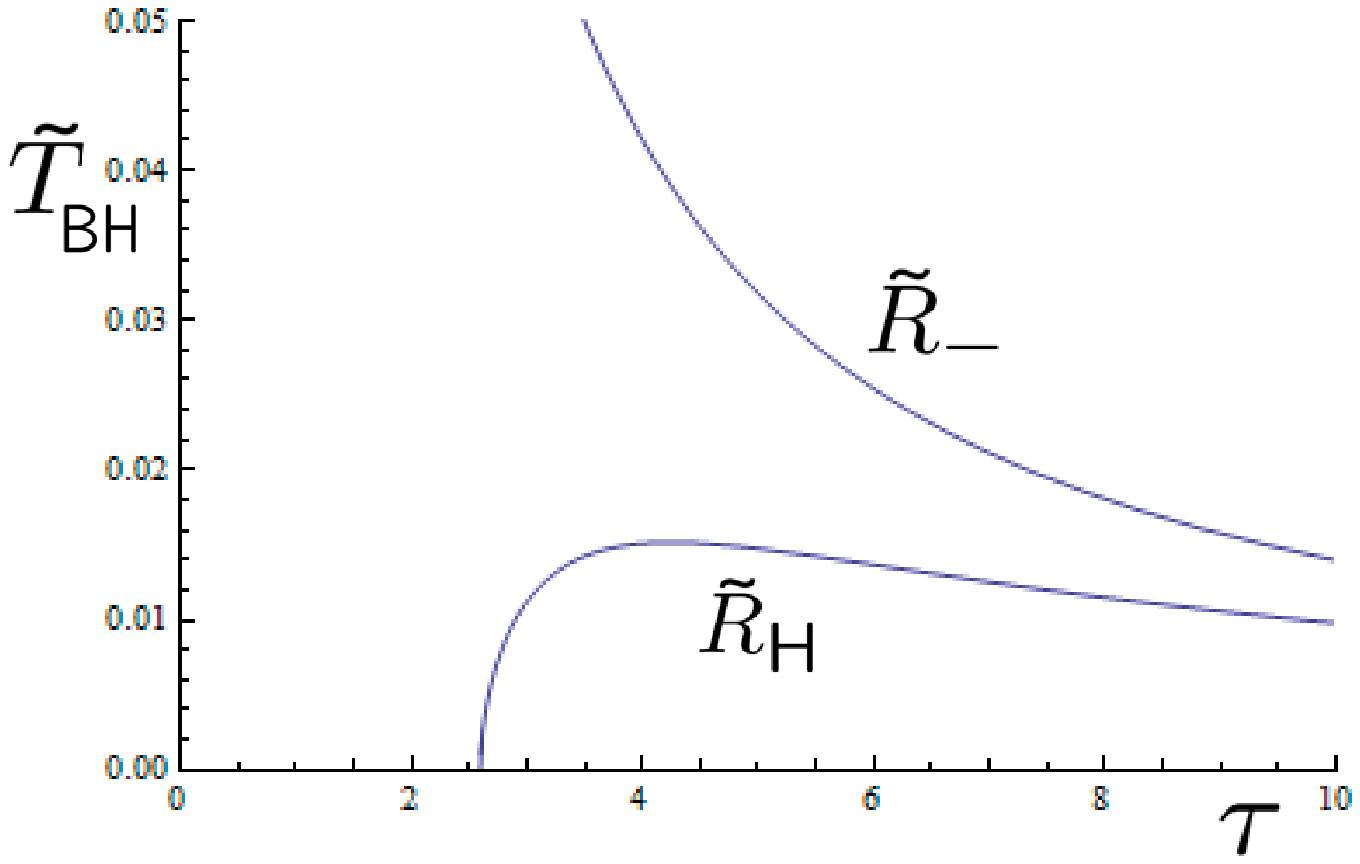}\\
\hskip 1.2cm (a) $\nt=1$
\hskip 1.8cm (b) $\nt=3$
\caption{
Black hole temperatures on two horizons ($\tilde R_-$
and $\tilde R_+$ or 
$\reh$) for $\nt=1$
and $3$. $\tilde T_{\rm {\scriptscriptstyle BH}}=Q
\, T_{\rm {\scriptscriptstyle BH}}$.}
\label{fig:temp}
\end{flushleft}
\vskip -3em
\end{figure}

Since the present spacetime is spherically symmetric
and the near horizon is ``static", we can calculate 
the surface gravity (see \cite{MN}
for  details).
Hence we find the black hole temperatures on the horizons
($T_{\rm {\scriptscriptstyle BH}}$) by the surface gravity
$\kappa_\pm$
as
\begin{eqnarray}
T_{\rm {\scriptscriptstyle BH}}^{(\pm)}={\kappa_\pm\over 2\pi}={\nt
\tilde R_\pm^{-{2\ns\over \nt}+1}
\over 16\pi \tau^2 Q}
\Bigl{|}
2\tau \tilde R_\pm^{{\ns\over \nt}-1}\mp \nt
\Bigr{|}
\,.
\end{eqnarray}
We depict the behaviour of the temperatures in Fig. 
\ref{fig:temp}.
They are finite and vanish in the limit of $\tau
\rightarrow \infty$, i.e., the extreme RN spacetime.

It may be interesting to discuss the thermodynamics
because we can define the entropy and temperature 
in these time-dependent spacetimes.
We can easily extend the present solution to arbitrary dimensions\cite{MN2}.
The details including the analysis of global structure 
and study of thermodynamical properties will be 
given elsewhere\cite{MN2}.
 Some questions for future work are :\\
(1) Can we find more realistic black hole solutions?
It may be straightforward to include rotation
 (cf \cite{Sh} ).
This  is under investigation.
As for neutral black holes in the Universe, it may be difficult to obtain
the analytic solution  because such a system is non-BPS state even in the 
static case, and the radius of
a black hole increases in time due to  accretion of matter .\\
(2) 
Can we extend black hole thermodynamics to time-dependent spacetimes?
The present time dependent solution may provide a good tool for analyzing
this question.\\
(3) 
Can we discuss some dynamical process with the present or extended solutions?
Black hole collision can be discuss in the contracting Universe ($t_0<0$) 
just as the KT spacetime.
We can also discuss the brane collisions  with multi time-dependent
branes, which is generalization of \cite{Gibbons:2005rt}.
\\
(4) Some solutions have a link to 
intersecting brane systems in higher-dimensional supergravity model.
If $\nt$ is a non-negative integer,
we may regard $\nt$ and $\ns$ as
numbers of branes.
It is true for $\nt=1$, in which case
we can derive the four-dimensional effective action (\ref{action})
from compactification of the time-dependent M2-M2-M5-M5 brane system in
11-dimensional supergravity theory (see Appendix in \cite{MN}).
Hence it may be interesting to see whether there is  any 
fundamental or deep reason for this link.\\
 Work along these lines is 
 in progress.


We would like to thank M.  Nozawa, N. 
Ohta, K. Shiraishi, and  K. Uzawa
for valuable  comments and discussions.
KM would acknowledge hospitality of 
DAMTP and the Centre for Theoretical Cosmology,
Cambridge University.
This work was partially supported 
by the JSPS Grants (No.19540308) and the
Japan-U.K. Joint Research Project.



\begin{thebibliography}{10}
\bibitem{TeV_BH}
T. Banks and W. Fischler, arXiv:hep-th/9906038;
S. Dimopoulos and G. Landsberg,
Phys. Rev. Lett. {\bf 87}, 161602 (2001);
S. B. Giddings and S. Thomas,
Phys. Rev. {\bf D65},  056010 (2002).



\bibitem{Hawking1974}
  S.W.~Hawking,
  Commun.\ Math.\ Phys.\  {\bf 43}, 199 (1975)
  [Erratum-ibid.\  {\bf 46}, 206 (1976)].



\bibitem{Strominger:1996sh}
  A.~Strominger and C.~Vafa,
  Phys.\ Lett.\  B {\bf 379}, 99 (1996)

\bibitem{Penrose} R. Penrose,
Nuovo Cimento {\bf 1}, 252 (1969).

\bibitem{Einstein:1945id}
  A.~Einstein and E.~G.~Straus,
  Rev.\ Mod.\ Phys.\  {\bf 17}, 120 (1945).
\bibitem{Mcvittie1933}
G. C.~McVittie,
Mon. Not. R. Astron. Soc.{\bf 93}, 325 (1933).
\bibitem{Nolan} 
B.C. Nolan,
Phys. Rev. D{\bf 58}, 064006 (1998);
Class. Quant. Grav. {\bf 16} 1227;3183  (1999);
\bibitem{SultanaDyer}
J. Sultana and C. C. Dyer, 
Gen.\ Rel.\ Grav. {\bf 37}, 1349 (2005).

\bibitem{MHC} 
H. Maeda, T. Harada, B.J. Carr, Phys. Rev. {\bf D77}, 024023 (2008).

\bibitem{CH} 
B.J. Carr, S.W. Hawking, Mon. Not. R. Astron. Soc.
{\bf 168}, 399 (1974).

\bibitem{accretion_BH_universe}
T. Jacobson, Phys. Rev. Lett. {\bf 83}, 2699 (1999); 
R. Bean and J. Magueijo, Phys. Rev. {\bf D66}, 063505 (2002);
A. Frolov and L. Kofman, JCAP {\bf 5}, 9 (2003); 
E. Babichev, V. Dokuchaev, Y. Eroshenko,
Phys. Rev. Lett. {\bf 93}, 021102 (2004).

\bibitem{Carter}
B. Carter, in {\it Black Holes}, edited
by C. DeWitt and J. DeWitt 
(Gordon and Breach, New York, 1973). 

\bibitem{SdS}
F. Kottler, Annalen Physik, {\bf 56}, 410 (1918).

\bibitem{BH} 
D.R. Brill and S.A. Hayward, 
Class. Quant. Grav. {\bf 11}, 359 (1994). 


\bibitem{KT}
D.~Kastor and J.H.~Traschen, 
Phys. Rev. D {\bf 47}, 5370  (1993).


\bibitem{BHKT}
  D.R.~Brill, G.T.~Horowitz, D.~Kastor and J.H.~Traschen,
  Phys.\ Rev.\  D {\bf 49}, 840 (1994).


\bibitem{Hartle:1972ya}
  J.B.~Hartle and S.W.~Hawking,
  Commun.\ Math.\ Phys.\  {\bf 26}, 87 (1972).

\bibitem{BC}
  K.~Behrndt and M.~Cvetic,
  Class.\ Quant.\ Grav.\  {\bf 20}  4177 (2003).

\bibitem{Sh}
  T.~Shiromizu,
  Prog.\ Theor.\ Phys.\  {\bf 102}, 1207 (1999).

\bibitem{MOU}
K.~Maeda, N.~Ohta and K.~Uzawa,
JHEP {\bf 0906},  051 (2009).

\bibitem{MN}
K. Maeda and M. Nozawa,
Phys. Rev. {\bf D81}, 044017
(2010).

\bibitem{power_law}
J.J. Halliwell, Phys. Lett. {\bf B 185}, 341 (1987).

\bibitem{footnote}
Similar time-dependent 
solution 
with one U(1) field
was found
[T. Maki and K. Shiraishi,
Class. Quant. Grav. {\bf 10}, 2171 (1993))].
But this solution is missing the part of $H_S$, which
is the reason why they failed to find a black hole. 

\bibitem{MN2}
K. Maeda and M. Nozawa,
in preparation.

\bibitem{HKS}
  S.~Hellerman, N.~Kaloper and L.~Susskind,
  JHEP {\bf 0106} (2001) 003.


\bibitem{Gibbons:2005rt}
  G.W.~Gibbons, H.~Lu and C.N.~Pope,
  Phys.\ Rev.\ Lett.\  {\bf 94}, 131602 (2005).




\end{thebibliography}
\end{document}